\documentclass{mn2e}
\usepackage{graphicx,amssymb,amsmath}

\newcommand{\beq}{\begin{equation}}
\newcommand{\enq}{\end{equation}}
\newcommand{\m}[1]{\boldsymbol{#1}}
\newcommand{\pa}{\partial}
\newcommand{\bra}[1]{\left#1}
\newcommand{\ket}[1]{\vphantom{\sqrt{0}}\right#1}
\newcommand{\mtext}[1]{\hspace{0.6cm}\mbox{#1}\hspace{0.6cm}}

\title[The force-free twisted magnetosphere]{The force-free twisted magnetosphere of a neutron star}
\author[T. Akg\"{u}n et al.]
{T.~Akg\"{u}n$^1$\thanks{E-mail: akgun@astro.cornell.edu}, J.A.~Miralles$^1$, J.A.~Pons$^1$ and P.~Cerd\'{a}--Dur\'{a}n$^2$
\\$^1$Departament de F\'{i}sica Aplicada, Universitat d'Alacant, Ap. Correus 99, 03080 Alacant, Spain
\\$^2$Departament d'Astronomia i Astrof\'{i}sica, Universitat de Val\`{e}ncia, Dr. Moliner 50, 46100, Burjassot, Val\`{e}ncia, Spain}

\begin{document}
\label{firstpage}
\pagerange{\pageref{firstpage}--\pageref{lastpage}}
\maketitle

\begin{abstract}
We present a detailed analysis of the properties of twisted, force-free magnetospheres of non-rotating neutron stars, which are of interest in the modelling of magnetar properties and evolution. In our models the magnetic field smoothly matches to a current-free (vacuum) solution at some large external radius, and they are specifically built to avoid pathological surface currents at any of the interfaces. By exploring a large range of parameters, we find a few remarkable general trends. We find that the total dipolar moment can be increased by up to $40\%$ with respect to a vacuum model with the same surface magnetic field, due to the contribution of magnetospheric currents to the global magnetic field. Thus, estimates of the surface magnetic field based on the large-scale dipolar braking torque are slightly overestimating the surface value by the same amount. Consistently, there is a moderate increase in the total energy of the model with respect to the vacuum solution of up to $25\%$, which would be the available energy budget in the event of a fast, global magnetospheric reorganization commonly associated with magnetar flares. We have also found the interesting result of the existence of a critical twist ($\varphi_{\rm max} \lesssim 1.5$ rad), beyond which we cannot find any more numerical solutions. Combining the models considered in this paper with the evolution of the interior of neutron stars will allow us to study the influence of the magnetosphere on the long-term magnetic, thermal, and rotational evolution.
\end{abstract}



\section{Introduction}
Soon after the discovery of pulsars, Goldreich \& Julian (1969) proposed the first realistic model of a neutron star magnetosphere in order to explain qualitatively the observations. In their model, a magnetic dipole aligned with the rotation axis of the star is able to fill the magnetosphere with plasma and produce a variety of interesting observational phenomena. Shortly afterwards, other models for rotating magnetospheres were constructed by Michel (1973a, 1973b, 1974). All these models are based on the assumption that the dynamics of the magnetosphere is dominated by the electromagnetic field, and the plasma pressure as well as its inertia are negligible. In such a case a reasonable approximation to the large-scale structure of the magnetosphere is given by force-free configurations, in which the electric and magnetic forces on the plasma are exactly balanced. For axially symmetric configurations, this condition leads to the so-called pulsar equation (Michel 1973b; Scharlemann \& Wagoner 1973), a partial differential equation for the {\it stream function} containing an additional unknown function that must be determined consistently by imposing continuity of the solution at the light cylinder. The first consistent solution to this equation with a dipole magnetic field near the star had to wait till the end of the 90s when Contopoulos, Kazanas \& Fendt (1999) were able to obtain a numerical solution by an iterative process. Since then, other authors have solved this equation confirming the validity of the solution (e.g. Timokhin 2006). More recently, solving the time-dependent equations of force-free electrodynamics (Komissarov 2002), numerical models for non-aligned magnetospheres of rotating neutron stars were obtained for the first time by Spitkovsky (2006), and since then other authors have obtained similar solutions (Kalapotharakos, Contopoulos \& Kazanas 2012; Kalapotharakos et al.\ 2012; P\'etri 2012; Li, Spitkovsky \& Tchekhovskoy 2012; Tchekhovskoy, Spitkovsky \& Li 2013; Philippov \& Spitkovsky 2014).

Although the force-free condition is a reasonable approximation for the global structure of the magnetosphere of a pulsar, it should be noted that it nevertheless is unlikely to be precisely satisfied everywhere in the magnetosphere of a neutron star, and there may be small regions (gaps) where particles are accelerated by the electric field along the magnetic field lines. Such processes are also necessary in order to explain emission mechanisms in the magnetosphere (Beloborodov \& Thompson 2007; Beloborodov 2013a, 2013b).

We will focus our attention on the class of neutron stars with the highest  magnetic field strength, $B\sim 10^{14}$ G, the so-called {\it magnetars}. The spectra of magnetars suggest the presence of a twist (a toroidal component) in the  magnetosphere (Tiengo et al.\ 2013; see Mereghetti, Pons \& Melatos 2015 for a review on magnetar properties). This twist may be maintained on long timescales by a transfer of helicity from the interior of the star (Thompson \& Duncan 1995), and also implies that the magnetosphere is not current-free. Thus, magnetosphere models are important in the context of long-term magnetic field evolution of neutron stars with strong magnetic fields (Vigan\`{o} et al.\ 2013). In the case of the twisted magnetosphere models that we discuss here, energy and helicity can be transferred from the stellar interior into the magnetosphere and vice versa, thus significantly affecting the evolution. Although rotation is crucial to explain the emission mechanism of ordinary pulsars, magnetars have a slow rotation rate and its effect will not be considered in this work. Thus, we will consider the force-free magnetosphere models of non-rotating neutron stars. Without rotation, the pulsar equation reduces to the standard Grad--Shafranov equation which determines the equilibrium structure of the magnetic field in a plasma. Although much simpler than the pulsar equation, the Grad--Shafranov equation contains an additional unknown function that cannot be determined by imposing continuity of the solution, and can be freely specified. Solutions to the Grad--Shafranov equation are of interest both in the astrophysical context of magnetic fields and in plasma physics in the context of magnetic confinement and fusion. Notwithstanding this great interest, analytic or semi-analytic solutions available for this case are rather limited (see, for example, Atanasiu et al.\ 2004 in the context of magnetic confinement and Thompson, Lyutikov \& Kulkarni 2002 in the context of magnetars), and, in general, numerical solutions are needed (Vigan\`{o}, Pons \& Miralles 2011; Glampedakis, Lander \& Andersson 2014; Fujisawa \& Kisaka 2014; Pili, Bucciantini \& Del Zanna 2015). Vigan\`{o} et al.\ (2011) use a magneto-frictional method to relax an initial (random) magnetic field to a force-free configuration. However, they require a surface current in order to connect their solution to a current-free field at the outer boundary. Glampedakis et al.\ (2014) solve for the interior and exterior equilibrium magnetic fields simultaneously applying the code of Lander \& Jones (2009) to the two regions, while implicitly forcing the magnetic field to remain finite at infinity. A similar approach is taken by Fujisawa \& Kisaka (2014), who in addition treat the core and crust separately by imposing magnetohydrostatic equilibrium in the core and Hall equilibrium in the crust, giving rise to surface currents at the crust-core interface. Pili et al.\ (2015) solve the Grad--Shafranov equation in a single domain encompassing the interior and the exterior by extending the interior solution to the low-density exterior.

In this paper we present axisymmetric non-relativistic force-free magnetosphere models for non-rotating stars with poloidal and toroidal magnetic fields. For this purpose, we obtain numerical solutions to the relevant Grad--Shafranov equation. Typical magnetar rotation periods lie in the range $2$ to $12$~s, which result in a light cylinder located at a distance $R_{\rm L} > 10^5$~km. In the region well inside the light cylinder (for $r \ll R_{\rm_L}$), the characteristic timescale to reach a force-free configuration, i.e.\ the Alfv\'{e}n crossing time (which is of the order of the light-crossing time $r/c$), is much smaller than the rotational period and rotation can be safely neglected. Only magnetic field lines from a small region around the magnetic pole extend to greater distances, connecting the neutron star to the light cylinder and beyond. For example, in the case of a dipolar magnetic field, the angle from the magnetic axis of a field line connected to the light cylinder is $\theta_{\rm L} \approx \sqrt{R_*/R_{\rm L}} < 0.01$~rad. In this work we adopt the simplifying assumption that near the poles, up to a certain critical field line, with $\theta > \theta_{\rm L}$, the magnetic field is current-free. We then assume that the force-free magnetic field with a toroidal component is confined within a region delimited by the critical current-free field line. This ensures that at large distances the magnetic field strength decreases sufficiently fast, approaching the current-free (vacuum) field, and eases the process of imposing boundary conditions.

The structure of this paper is as follows: section \S\ref{section_review} is a review of some relevant background theory related to the problem, and that is useful for the magnetosphere model, which is then presented in \S\ref{section_model}. The numerical methods applied are briefly described in \S\ref{section_numerical}, sample results are discussed in \S\ref{section_results}, and the conclusions are presented in \S\ref{section_conclusions}.


\section{Relevant equations and notation}\label{section_review}
\subsection{The Grad--Shafranov equation}
In general, any magnetic field (or more generally, any divergenceless, i.e.\ \emph{solenoidal} field) can be written as the sum of a \emph{poloidal} and a \emph{toroidal} field (Chandrasekhar 1981). In particular, in the case of axisymmetry, defining the poloidal and toroidal functions as $P$ and $T$, respectively, the magnetic field can be expressed as
	\beq
	\begin{split}
	\m{B} & = \m\nabla P \times \m\nabla\phi + T \m\nabla\phi \\
	& = - \frac{\pa_z P}{\varpi} \m{\hat\varpi} + \frac{\pa_\varpi P}{\varpi} \m{\hat{z}}
	+ \frac{T}{\varpi} \m{\hat\phi} \\
	& = \frac{\pa_\theta P}{r^2\sin\theta} \m{\hat{r}}
	- \frac{\pa_r P}{r\sin\theta} \m{\hat\theta} + \frac{T}{r\sin\theta} \m{\hat\phi} \ .
	\end{split}
	\label{mag_PT}
	\enq
$P$ and $T$ are (\emph{stream}) functions of radius $r$ and polar angle $\theta$ in spherical coordinates $(r,\theta,\phi)$ and are functions of cylindrical radius $\varpi$ and $z$ in cylindrical coordinates $(\varpi,\phi,z)$. Here, the gradient of the azimuthal angle $\phi$ is used for mathematical convenience, and is related to the azimuthal unit vector through $\nabla\phi = \m{\hat\phi}/\varpi = \m{\hat\phi}/r\sin\theta$.

The magnetic field can alternatively be expressed in terms of the vector potential as
	\beq
	\begin{split}
	\m{B} = \m\nabla \times \m{A} \ .
	\end{split}
	\label{mag_A}
	\enq
The vector potential is undetermined up to a gauge freedom $\m{A} \to \m{A} + \m\nabla\psi$. For an axisymmetric field this implies that the radial ($A_r$) and colatitudinal ($A_\theta$) components of the vector potential are undetermined up to some function. Comparing equations (\ref{mag_PT}) and (\ref{mag_A}), note that the poloidal function is related to the azimuthal component of the vector potential, while the toroidal function depends on a combination of the remaining two components,
	\beq
	\begin{split}
	P & = \varpi  A_\phi = r \sin\theta A_\phi \  , \\
	T & = \varpi B_\phi = r\sin\theta B_\phi 
	= \bra{[} \pa_r (r A_\theta) - \pa_\theta A_r \ket{]}\sin\theta \ .
	\end{split}
	\enq
While $A_r$ and $A_\theta$ are individually undetermined up to a gauge freedom, their combination giving the function $T$ is determined.

The current density also has corresponding poloidal and toroidal components,
	\beq
	\frac{4\pi\m{J}}{c} = \m\nabla\times\m{B}
	= - \triangle_{\rm GS} P \m\nabla\phi + \m\nabla T \times \m\nabla\phi \ ,
	\label{mag_J}
	\enq
where $\triangle_{\rm GS}$ is the so-called Grad--Shafranov operator,
	\beq
	\begin{split}
	\triangle_{\rm GS} & = \varpi^2\m\nabla\cdot(\varpi^{-2}\m\nabla) \\
	& = \nabla^2 - \frac{2}{\varpi} \pa_\varpi
	= \pa_\varpi^2 - \frac{1}{\varpi} \pa_\varpi + \pa_z^2 \\
	& = \pa_r^2 + \frac{\sin\theta}{r^2} \pa_\theta
	\left(\frac{\pa_\theta}{\sin\theta}\right)
	= \pa_r^2 + \frac{1-\mu^2}{r^2} \pa_\mu^2 \ .
	\end{split}
	\label{GS_op}
	\enq
We will also find it convenient to occasionally use the notation $\mu \equiv \cos\theta$. Note that the poloidal magnetic field is due to the toroidal current density, and the toroidal magnetic field is due to the poloidal current density.

For static axisymmetric equilibria in fluids, the magnetic force cannot have an azimuthal ($\m{\hat\phi}$) component as there is no corresponding hydrostatic force that can act to balance it. This requirement can be expressed as $\m\nabla P \times \m\nabla T = 0$, which implies that the poloidal and toroidal functions must be functions of one another, for example $T = T(P)$. Note that this includes the special cases when the toroidal function is constant or zero, or when the poloidal field is zero. Then the force density, $\m{f}$, reduces to
	\beq
	4\pi\m{f} = (\m\nabla\times\m{B})\times\m{B}
	= - \frac{\triangle_{\rm GS} P + G(P)}{\varpi^2} \m\nabla P \ ,
	\label{lorentz}
	\enq
where we define $G(P) = T(P) T'(P)$. In a \emph{barotropic} fluid, the force density per unit mass $\m{f}/\varrho$, where $\varrho$ is the density, must further be expressible as a gradient of a potential. In fact, this is true in general, even in the context of more general magnetic forces, both in normal and type II superconducting fluids (Akg\"{u}n \& Wasserman 2008). This then gives the so-called Grad--Shafranov equation\footnote{Arguably, it may be fairer to name the equation after L\"{u}st and Schl\"{u}ter, as their paper seems to precede those of either Grad or Shafranov.}
	\beq
	\triangle_{\rm GS} P + G(P) = \varrho \varpi^2 F(P) \ ,
	\label{GS_eq}
	\enq
where $F(P)$ is some arbitrary function of $P$ (L\"{u}st \& Schl\"{u}ter 1954; Chandrasekhar 1956a, 1956b; Chandrasekhar \& Prendergast 1956; Prendergast 1956; Shafranov 1957, 1958, 1966; Grad \& Rubin 1958). In particular, observe that:
	\begin{itemize}
	\item[(a)] force-free fields are given by the equation $\triangle_{\rm GS} P + G(P) = 0$, i.e.\ $F(P) = 0$;
	\item[(b)] current-free (vacuum) fields are further restricted by the individual requirements $\triangle_{\rm GS} P = 0$ and $G(P) = 0$.
	\end{itemize}

Also note that the current density in a force-free field is given by (from equation \ref{mag_J})
	\beq
	\frac{4\pi\m{J}}{c} = T'(P) \m{B} \ .
	\label{mag_Jff}
	\enq
Thus, the current is parallel to the magnetic field (i.e.\ $\m{B}$ is a Beltrami vector field), and the current flows along the magnetic field lines, which lie on the magnetic surfaces defined by constant $P$.

The Grad--Shafranov equation is of major interest in plasma physics as well as in astrophysics, however only a limited range of analytical solutions are available (for a review, see Atanasiu et al.\ 2004). Numerical solutions have been constructed in the context of magnetic equilibria in barotropic fluid stars with rotation (Tomimura \& Eriguchi 2005; Yoshida \& Eriguchi 2006; Yoshida, Yoshida \& Eriguchi 2006; Lander \& Jones 2009), without rotation (Armaza, Reisenegger \& Valdivia 2015), and including general relativistic effects (Ioka \& Sasaki 2003; Ciolfi et al.\ 2009; Ciolfi, Ferrari \& Gualtieri 2010). The force-free equation has also been applied to the magnetospheres of pulsars and magnetars (Thompson et al.\ 2002; Spitkovsky 2006; Beskin 2010; Vigan\`{o} et al.\ 2011; Glampedakis et al.\ 2014; Fujisawa \& Kisaka 2014; Pili et al.\ 2015). 

Incidentally, in the context of magnetic field evolution, Hall-inactive (or Hall equilibrium) magnetic fields also satisfy a Grad--Shafranov equation, with the only difference being that the mass density $\varrho$ is replaced by the electron number density $n_e$ in the source term on the right hand side of equation (\ref{GS_eq}) (see, for example, Gourgouliatos et al.\ 2013; Marchant et al.\ 2014). Moreover, the force-free solutions for linear $G(P)$ are of the same form as the Ohmic modes (Marchant et al.\ 2014).

Finally, it is worth noting that the magnetic field and current density are related to the magnetic flux $\Phi$ and current $I$ through
	\beq
	\Phi = \int \m{B}\cdot d\m{S} \mtext{and}
	I = \int \m{J}\cdot d\m{S} \ .
	\enq
The poloidal and toroidal functions $P$ and $T$ are constant on the same magnetic surfaces, and the above definitions imply that the poloidal function $P$ corresponds to the flux passing through the area enclosed by the corresponding magnetic surface, and $T$ corresponds to the current through the same area. More precisely, carrying out the integrations over equatorial circles delineated by the magnetic surfaces (and thus having unit normal vectors $\m{\hat{z}}$), we get $\Phi = 2\pi P$ and $I = c T / 2$.

\subsection{Auxiliary definitions}\label{section_auxiliary}
In order to characterize the different models of magnetospheres, it is useful to define several quantities of interest as described next.

\subsubsection{Magnetic energy and helicity}\label{section_energy}
In this work, we will be concerned with force-free magnetic fields. In general, the energy of such fields can be expressed entirely in terms of surface integrals (Chandrasekhar 1981)
	\beq
	8\pi {E} = \int B^2 \, dV
	= \oint B^2 (\m{r}\cdot d\m{S}) - 2 \oint (\m{r}\cdot \m{B}) (\m{B}\cdot d\m{S}) \ .
	\label{energy}
	\enq
Note that the second term vanishes over magnetic surfaces where $\m{B} \perp d\m{S}$. The derivation of this important formula is given in Appendix \ref{section_proof}.

Magnetic helicity is defined as (Berger \& Field 1984; Berger 1999, and references therein)
	\beq
	{H} = \int \m{A}\cdot\m{B} \, dV \ ,
	\enq
and measures the degree to which the magnetic field wraps around itself, and is related to the \emph{linking number} in topology. Under certain conditions (including ideal magnetohydrodynamics) helicity is conserved. In terms of the components of the magnetic field $\m{B}$ and the vector potential $\m{A}$ defined in equations (\ref{mag_PT}) and (\ref{mag_A}), and carrying out some integrations by part, the helicity can be written as
	\beq
	{H} = 2 \int A_\phi B_\phi \, dV
	- \oint A_\theta A_\phi (\m{\hat{r}} \cdot d\m{S}) \ .
	\label{helicity}
	\enq
In obtaining this result we have made use of the fact that $P=0$ along the axis. The remaining surface integral can be made to vanish through an appropriate choice of the gauge, namely that $A_\theta = 0$ at the surface, and therefore we will not be concerned with it in what follows.

\subsubsection{Twist}\label{section_twist}
A quantity closely related to helicity is the twist, which we define as the azimuthal extent of a field line (measured in radians). Clearly, in the absence of a toroidal field the twist is zero, and it increases with toroidal field strength and field line length. The twist can be calculated using the defining equations for a field line
	\beq
	\frac{dr}{B_r} = \frac{r d\theta}{B_\theta}
	= \frac{r\sin\theta d\phi}{B_\phi} = \frac{d\ell}{B_{\rm pol}}\ ,
	\enq
where $d\ell$ is the poloidal field line element (obtained through the projection of the field line onto the ($r$,$\theta$) plane), $B_{\rm pol}$ is the poloidal field magnitude ($B_{\rm pol} = \sqrt{B_r^2 + B_\theta^2}$), and $B_\phi$ is the toroidal field magnitude. From the last two equations, it follows that the twist is given by
	\beq
	\varphi \equiv \Delta\phi =
	\int_{0}^{\ell} \frac{B_\phi}{B_{\rm pol} r\sin\theta} d\ell \ ,
	\label{twist}
	\enq
where $\ell$ is the total length of the poloidal field line, and all quantities in the integration are evaluated along this line.

\subsubsection{Multipole content}
We define the multipole strength normalized to the surface as,
	\beq
	a_l \equiv \frac{r^l A_l(r)}{R_\star^l} \ ,
	\label{multipole_content}
	\enq
where $A_l(r)$ is the $l$th component of the multipole expansion of the poloidal function $P(r,\theta)$ at some radius $r$ (cf.\ Appendix \ref{section_multipole}). The multipole content is constant for a current-free field, while it will vary with radius for a force-free field with a twist. Thus, the multipole expansion at some radius beyond the largest extent of the currents (present in the toroidal region) serves as a measure of the deviation from the field at the stellar surface due to those currents.

\subsection{Dimensions}\label{section_dimensions}
Throughout this work, we will express all quantities in dimensionless units. All lengths will be measured in stellar radii ($R_\star$) and the magnetic field strength will be measured in units of some $B_{\rm o}$. We choose the normalization in such a way that for a dipolar poloidal function the magnetic field at the pole ($r=R_\star$ and $\theta = 0$) becomes $B_r = 2 B_{\rm o}$. In addition, in the purely dipolar case $B_{\rm o}$ corresponds to the surface magnetic field strength at the equator ($r=R_\star$ and $\theta = \pi/2$). All other dimensions used in the paper can be derived from these two definitions. The most important ones are listed in Table \ref{table_dimensions}.

\begin{table}
\caption{List of relevant quantities, notation and units.}
\label{table_dimensions}
\center
\begin{tabular}{lll}
\hline \hline
Quantity & Notation & Units \\
\hline \hline
Radius & $r$ & $R_\star$ \\
Toroidal function & $T$ & $B_{\rm o} R_\star$ \\
Poloidal function & $P$ & $B_{\rm o} R_\star^2$ \\
Magnetic field strength & $B$ & $B_{\rm o}$ \\
Energy & ${E}$ & $B_{\rm o}^2 R_\star^3$ \\
Helicity & ${H}$ & $B_{\rm o}^2 R_\star^4$ \\
Twist & $\varphi$ & ${\rm rad}$ \\
\hline \hline
\end{tabular}
\end{table}

\section{Force-free magnetosphere with a toroidal field}\label{section_model}
\begin{figure*}
	\centerline{\includegraphics[width=1\textwidth]{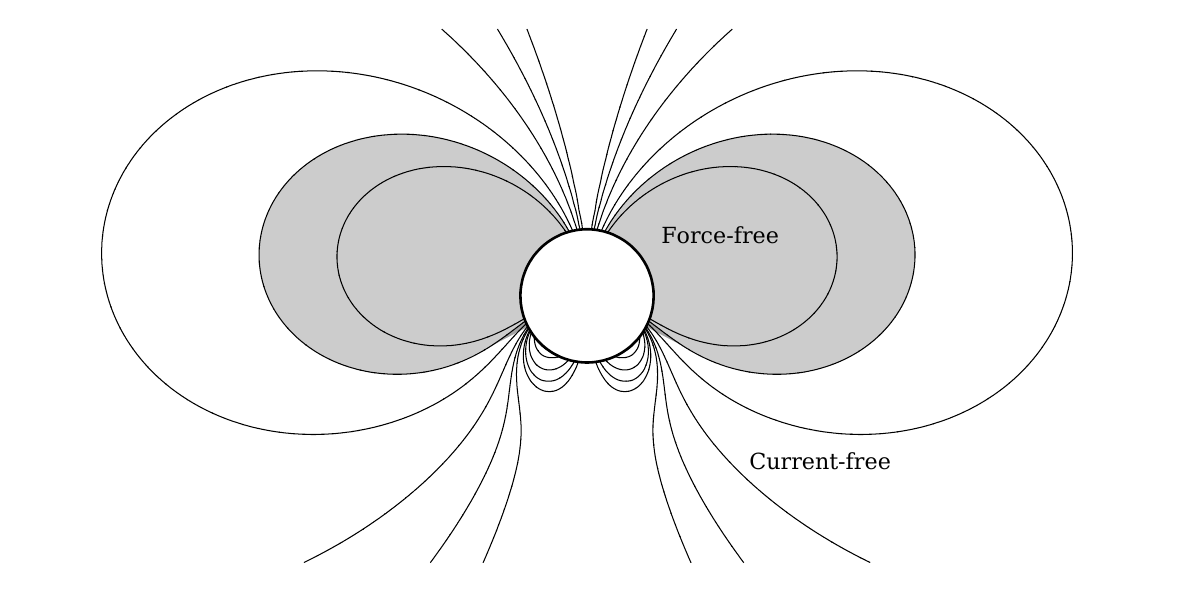}}
	\caption{An illustration of the magnetic field structure considered in this paper. The star is shown as a white circle, the force-free region containing the toroidal field is shown in gray, and the surrounding current-free (purely poloidal) region is shown in white. A combination of a dipolar and a quadrupolar component is depicted in the figure, and as a consequence the magnetic field lines are not symmetric with respect to the equator. In a realistic numerical solution the number of multipoles is arbitrary, and the resulting structure is somewhat different than shown here.}
	\label{fig_geometry}
	\end{figure*}

In this work, we construct a force-free magnetosphere with a toroidal field confined to a region defined by a certain poloidal field line. Within the toroidal region there are currents, while outside of it the magnetic field is current-free. The geometry of a sample magnetic configuration of this kind is illustrated in Figure \ref{fig_geometry}. We assume that the toroidal function is given in terms of the poloidal function through
	\beq
	T(P) = \left\{
	\begin{aligned}
	& s(P-P_{\rm c})^\sigma & \mtext{for} P \geqslant P_{\rm c} \ , \\
	& 0 & \mtext{for} P < P_{\rm c} \ .
	\end{aligned}
	\right.
	\label{T_of_P}
	\enq
In terms of the Heaviside step function
	\beq
	\Theta(x) = \left\{
	\begin{aligned}
	& 1 & \mtext{for} x \geqslant 0 \ , \\
	& 0 & \mtext{for} x < 0 \ ,
	\end{aligned}
	\right.
	\label{heaviside}
	\enq
we can express the toroidal function as\footnote{More precisely, the toroidal function can be written in terms of the ramp function, which is defined as $R(x)=x \Theta(x)$. The ramp ($R$), Heaviside ($\Theta$) and Dirac delta ($\delta$) functions are related through $R'(x) = \Theta(x)$ and $\Theta'(x) = \delta(x)$. Additionally, note that the Dirac delta function satisfies $x\delta(x) = 0$. This property is significant as it ensures that the derivative of the toroidal function for the $\sigma=1$ case is still only a step function and not a (problematic) delta function.}
	\beq
	T(P)=s(P-P_{\rm c})^\sigma \Theta(P-P_{\rm c}) \ .
	\enq
In order to avoid divergences in the current density we must have $\sigma \geqslant 1$. The magnetic field configuration is described by the Grad--Shafranov equation (equation \ref{GS_eq}), which in this case becomes
	\beq
	\triangle_{\rm GS} P + \sigma s^2 (P - P_{\rm c})^{2\sigma-1} \Theta(P-P_{\rm c}) = 0 \ .
	\label{GSeqff}
	\enq
Thus, there are three important parameters that define the toroidal field: the coefficient $s$ which determines the relative strength of the toroidal field with respect to the poloidal field; the critical field line $P_{\rm c}$ which defines the size of the toroidal region (in the magnetic coordinate $P$); and the power index $\sigma$ which sets the functional dependence between the toroidal and poloidal fields.

In particular, for $\sigma = 1$, the Grad--Shafranov equation becomes linear. In this case, the homogeneous part is of the same form as in Ciro \& Caldas (2014), so, in principle, the same analytical solutions can be used for the toroidal region. These should then be matched to the vacuum solutions outside the toroidal region, which, in general, will contain any number of unknown multipoles. Thus, analytic solutions involve intractable infinite sums of multipoles, and instead we seek numerical solutions satisfying the requirement that beyond the region containing the currents, the field (continuously) matches to a vacuum solution that vanishes at infinity.

Since the Grad--Shafranov equation for $\sigma=1$ is an elliptic partial differential equation, the solutions are guaranteed to be smooth (continuous and differentiable) to a certain degree. In particular, the poloidal function should be at least twice differentiable. This implies that the magnetic field, which involves the first derivative of the poloidal function (equation \ref{mag_PT}), is continuous throughout the entire region, and, in particular, across the boundary of the toroidal region. On the other hand, while the Lorentz force (equation \ref{lorentz}) is guaranteed to be zero everywhere and therefore continuous in such a configuration, the current density (equation \ref{mag_J}) has a discontinuity on the magnetic surface enclosing the toroidal field in the form of a step function, which arises as a consequence of the discontinuity in the first derivative of the toroidal function. At this boundary, the toroidal (azimuthal) part of the current density ($- \triangle_{\rm GS} P \m\nabla\phi$) vanishes, and the poloidal part of the current density ($\m\nabla T \times \m\nabla\phi$) is parallel to the poloidal magnetic field ($\m\nabla P \times \m\nabla\phi$) which defines the boundary, implying that the currents flow \emph{on}, but not \emph{out} of the enclosing magnetic surface. Nevertheless, since the magnetic field is continuous, crucially, there are no surface currents. To reemphasize, this ``current at a surface'' is a discontinuity in the form of a step function $\Theta(P - P_{\rm c})$ without any further undesirable effects on the physical quantities of interest, and is not to be confused with a ``surface current'' which (in addition to being in a different direction) is a severe pathology in the form of a delta function $\delta(P - P_{\rm c})$ causing a discontinuity in the magnetic field.

In order to avoid a discontinuity in the current altogether, a higher power relation must be taken for the toroidal function ($\sigma > 1$). In that case, the differential equation becomes non-linear and analytic solutions are no longer available. We observe that increasing the power $\sigma$ concentrates the toroidal field near the equator, and thus reduces its effect on the outlying poloidal field. This is the justification for taking $\sigma$ to be small but larger than 1, with the usual choice being $\sigma=1.1$ (Lander \& Jones 2009). However, we reiterate that the limiting case of $\sigma=1$ is also perfectly well-behaved for all our purposes.

More generally, the Grad--Shafranov equation is a second-order non-linear inhomogeneous partial differential equation, and the existence and uniqueness of its solutions are not trivial matters. To be precise, in our case, equation (\ref{GSeqff}) is quasi-linear, since it is linear in the second (highest) derivatives. It can be written in the form $\Delta_{\rm GS} P = f(P)$. If $f'(P) \geqslant 0$, it is possible to use a maximum principle to prove local uniqueness of the solution (see Taylor 1996, chapter 14). However, this is not the case: since $\sigma \geqslant 1$, for any value of $P$ we have $f'(P) \leqslant 0$. Therefore, uniqueness of the solution of the Grad--Shafranov equation cannot be guaranteed in general. For sufficiently small values of $T'(P)$, Bineau (1972) proved the uniqueness of force-free solutions, provided the solution domain is bounded and the field is not vanishing anywhere. Notwithstanding these complications, we were able to construct numerical solutions for a wide range of parameters without significant difficulty (up to some maximum value of $s$, as discussed later on).

The overall strength of the magnetic field scales out in our calculations and our results do not explicitly depend on it. On the other hand, the \emph{relative} strength of the toroidal and poloidal fields is determined (non-linearly) by the parameters of the toroidal function $s$, $P_{\rm c}$ and $\sigma$. In the units listed in Table \ref{table_dimensions}, the parameter $s$ has dimensions of $B_{\rm o}^{1-\sigma} R_\star^{1-2\sigma}$, and the toroidal field is then given in units of $B_{\rm o}$.

\section{Numerical methods}\label{section_numerical}
\subsection{Numerical solution of the Grad--Shafranov equation}
We have directly solved the (axisymmetric) Grad--Shafranov equation (equation \ref{GSeqff}) for a force-free magnetic field numerically by discretizing the equation using a uniform grid in radius and polar angle and imposing boundary conditions at the stellar surface, along the axes, and at some arbitrary external radius where the field is current-free. For a toroidal function $T$ of the form given by equation (\ref{T_of_P}), the discretized equations for the poloidal function $P$ form, in general, an algebraic non-linear system of equations that can be expressed as a block tridiagonal system with a non-linear source term that depends on $P$ implicitly through the function $G(P)$. A solution can be found providing an initial guess for $P$ for the non-linear term, then solving the linear algebraic system of equations, and repeating the process iteratively until convergence. An advantage of the numerical method is that it can deal with non-linear functions for $T(P)$, such as the step function considered here.

We write the Grad--Shafranov equation (equation \ref{GSeqff}) in discrete form through
	\beq
	\begin{split}
	\frac{P_{i+1,j} - 2P_{i,j} + P_{i-1,j}}{(\Delta r)^2}
	- \frac{P_{i,j+1} - P_{i,j-1}}{2 r_i^2 \Delta\theta} \cot\theta_j \\
	+ \frac{P_{i,j+1} - 2P_{i,j} + P_{i,j-1}}{r_i^2 (\Delta\theta)^2} 
	= - G(P_{i,j}^{\rm old}) \ ,
	\end{split}
	\label{GSeqff_discrete1}
	\enq
where the indices $i$ and $j$ correspond to the grid points $(r_i,\theta_j)$. The source term on the right-hand side is given in terms of the previous (old) guess for $P_{i,j}$ through
	\beq
	G(P_{i,j}^{\rm old}) = \sigma s^2 (P_{i,j}^{\rm old} - P_{\rm c})^{2\sigma-1} 
	\Theta(P_{i,j}^{\rm old}-P_{\rm c}) \ .
	\label{GSeqff_discrete2}
	\enq

We impose Dirichlet boundary conditions along the axis (by setting $P=0$) and at the stellar surface, where the form of the function $P$ is to be determined by the interior magnetic field.  At the external radius, we require that the field smoothly 
match to a vacuum field solution by imposing Neumann boundary conditions on the derivative of $P$. This is accomplished by carrying out a multipole expansion at some radius beyond the largest extent of the toroidal region, and imposing that each multipole decay radially, consistently with its corresponding vacuum profile. This requirement can also be implemented in other equivalent ways, and each has been found to work excellently.

For a given set of parameters $s$, $P_{\rm c}$ and $\sigma$, we solve the resulting block tridiagonal system by standard methods based on the tridiagonal matrix algorithm, also known as the Thomas algorithm (Thomas 1949), to determine the updated (new) guess for $P_{i,j}$. When a non-linear toroidal field (such as the step function considered here) is present, we need to carry out iterations, as the shape of the toroidal region is not known beforehand, and must be calculated consistently. At each iteration we calculate the square of the difference between the previous guess and the updated solution, averaged over the entire grid, and check for convergence of the solution. Thus, we define the correction to the previous guess at the $k$th iteration as
	\beq
	\chi_k^2 \equiv \sum_{i,j}^{N_r,N_\theta}
	\frac{(P_{i,j}^k - P_{i,j}^{k-1})^2}{N_r N_\theta} \ .
	\enq
Here the summation is carried out over the entire (two dimensional) grid of $N_r \times N_\theta$ points, and $P_{i,j}^0$ is the first starting guess. We consider that convergence is achieved once the value of $\chi_k^2$ is sufficiently small, typically many orders of magnitude less than $10^{-6}$, but we accept solutions with corrections up to that level. Once convergence is achieved, we calculate several quantities of interest, among them energy, helicity and maximum twist. We also study the dependence of these quantities on the parameters of the toroidal field $s$, $P_{\rm c}$, and $\sigma$.

Throughout the iterations we maintain the three parameters $s$, $P_{\rm c}$ and $\sigma$ fixed. This is in contrast to the iteration scheme of Pili et al.\ (2015), who instead require the critical field line containing the currents to pass through a given point on the equatorial plane, and therefore allow for $P_{\rm c}$ to change between iterations. This subtle difference in the iteration schemes may result in convergence to different results when there are multiple solutions for the same parameters (since the Grad--Shafranov equation may not have unique solutions, as discussed in \S\ref{section_model}), and may explain some of the differences between the two works.

As will be discussed in greater detail in \S\ref{section_accuracy}, we have performed a number of tests on accuracy. We have confirmed that the code is able to reproduce the analytic cases for a purely poloidal field ($s=0$), as well as the analytic solutions for the linear case $T=sP$ (with $s \ne 0$). For the latter case, the solution is given in terms of spherical Bessel functions (as discussed, for example, in Atanasiu et al.\ 2004 and Vigan\`{o} et al.\ 2011), and we have adopted analytical boundary conditions, since these solutions cannot be matched to a vacuum.

	\begin{figure*}
	\centerline{\includegraphics[width=1\textwidth]{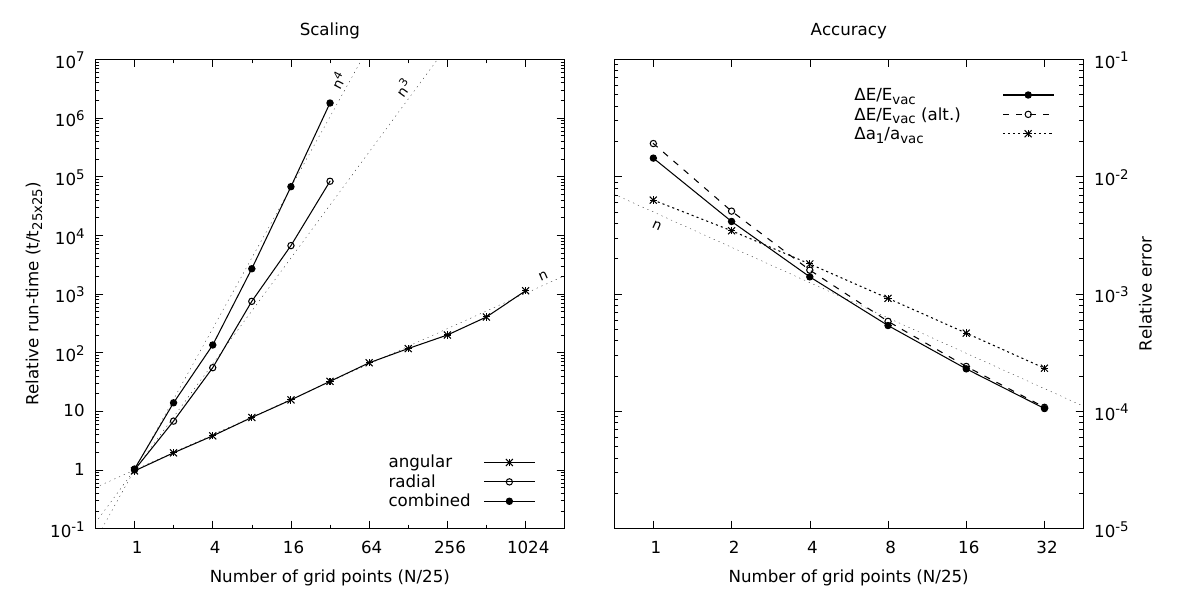}}
	\caption{Scaling of run-time (left) and accuracy (right) with the number of grid points. {\bf Left:} the scaling of our code with the number of grid points is shown on a log--log plot. The run-time ($t$) is rescaled by the first experiment's execution time ($t_{25\times 25}$), and the number of grid points ($N$) is shown in units of 25 ($n=N/25$). Scaling relations for the angular, radial and combined (both angular and radial) grids are shown. Power-law relations approximating each of the scaling tests are shown in dotted lines. {\bf Right:} sample tests of accuracy as functions of the number of grid points are shown on a log--log plot. The numerically calculated energy $E$ and dipole strength $a_1$ are shown relative to their respective vacuum values for the purely poloidal case ($s=0$). As discussed in \S\ref{section_energy2}, the energy can be calculated either as a volume integral plus a surface integral at some outer radius (solid line), or directly as a surface integral at the stellar surface (dashed line).}
	\label{fig_scaling}
	\end{figure*}

\subsection{Numerical computation of the energy}\label{section_energy2}
The energy of a force-free magnetic field can be calculated in terms of surface integrals through the formula given by equation (\ref{energy}) as long as the magnetic field is differentiable (which then implies that it is also continuous). This requirement, in turn, implies that the poloidal function $P$ should be twice differentiable and the toroidal function $T$ should be once differentiable (Appendix \ref{section_proof}), which are satisfied for all $\sigma \geqslant 1$ (and, in particular, are satisfied in the limit $\sigma \to 1$).

Thus, we can calculate the energy stored in the entire magnetosphere (from the stellar surface to infinity) using the value of the magnetic field (as determined through the functions $P$ and $T$) specified at the stellar surface. This provides an alternative form of checking for the continuity of the magnetic field, since the energy can also be calculated over a finite volume from the stellar surface up to an arbitrary radius extending beyond the toroidal region and where the magnetic field is that of a vacuum, plus a surface integration at that radius for the vacuum field extending to infinity. If these two energies are not consistent, then there may be surface currents, and consequently, magnetic field discontinuities in some regions. As shown in Figure \ref{fig_scaling} and discussed in \S\ref{section_accuracy}, the energies calculated in these two ways are indeed in good agreement.

It is worth noting that although the energy can be written purely as a surface integration, this surface integration (equation \ref{energy}) involves the components of the magnetic field $\m{B}$, in particular $B_\theta$, and therefore involves radial derivatives of the poloidal function $P$, which are only determined numerically. Explicitly, from equation (\ref{energy}) it follows that the energy contained in the volume beyond a radius $r$ can be written as
	\beq
	8\pi{E} = \int (B_r^2 - B_\theta^2 - B_\phi^2) r dS_r \ ,
	\enq
noting that the surface normal vector at $r$ is $\m{\hat{n}} = - \m{\hat{r}}$. While $B_\phi$ is given analytically at the surface (through the function $T$), and $B_r$ can (in principle) be constructed analytically from the function $P$ (through its angular derivatives), $B_\theta$ involves radial derivatives of $P$ and always needs to be determined numerically. Moreover, we calculate the first radial derivatives using \emph{forward} differences, which are less precise than \emph{central} differences used in the interior of the radial grid.

\subsection{Scaling}\label{section_scaling}
We carry out three scaling experiments to determine the run-time as a function of the number of grid points. The resulting scaling is shown on a log--log plot on the left-hand side in Figure \ref{fig_scaling}. Starting with a $25\times 25$ grid in radius and angle, three scaling tests are performed: first, the angular grid is increased in multiples of two, while the radial grid is kept constant (denoted as \emph{angular}); next, starting from the same initial configuration, the same is carried out for the radial grid, while the angular grid is kept constant (denoted as \emph{radial}); and finally, both the radial and the angular grid are simultaneously doubled (denoted as \emph{combined}). The execution time (run-time) depends both on machine specifications, and on the machine load at the time the test is carried out. A typical run of 30 iterations for a given $s$ and $P_{\rm c}$ on a $100\times 100$ grid takes about $\sim 10$ seconds on a desktop computer. As fluctuations in the run-time can at times be significant, we choose to express instead the \emph{relative} run-time, in units of that of the starting experiment performed on a $25\times 25$ grid (denoted as $t_{25\times 25}$ in the figure). Power-law relations approximating each of the scaling tests are shown in dotted lines. Our numerical procedure depends on the order in which the indices are implemented. In our case the first index is the radial index, and the CPU time scales as $N_r^3$, while it scales linearly with the angular index $N_\theta$. These scaling relations should be kept in mind when implementing such numerical methods, and the better scaling index should be chosen for the higher number of grid points. Overall, increasing the accuracy requires both the radial and angular grid to be increased. In this case, the scaling of the code with the number of grid points ($N\equiv N_r N_\theta$) is $N_r^3 N_\theta$.

\subsection{Accuracy}\label{section_accuracy}
We also perform checks on the accuracy of the code as a function of the number of grid points. Sample results are shown on a log--log plot on the right-hand side in Figure \ref{fig_scaling}. Imposing a dipolar field at the surface, and starting with a grid size of $25\times 25$, we repeatedly double the grid in each dimension (thus quadrupling the total number of points) and compare the numerical results for the energy and the dipole strength for the current-free (purely poloidal) case (corresponding to $s=0$) with their exact values. As noted in \S\ref{section_energy2}, the energy can be calculated either as a volume integral plus a surface integral at some outer radius (solid line), or directly as a surface integral at the stellar surface (dashed line). We express the relative difference between these two numerical results and the exact vacuum value (which for a dipole is ${E}_{\rm vac} = 1/3$, cf.\ equation \ref{energy_vacuum2}, in the units listed in Table \ref{table_dimensions}) as
	\beq
	\frac{\Delta {E}}{{E}_{\rm vac}} =
	\frac{{E} - {E}_{\rm vac}}{{E}_{\rm vac}} \ .
	\label{energy_relative}
	\enq
The volume plus surface integration (solid line) appears to be slightly more precise than the purely surface integration (dashed line). This is a consequence of the fact that the radial derivatives used in the surface integration have lower precision than the derivatives used in the volume integration (as noted in \S\ref{section_energy2}).

Similarly, we compare the dipole strength as defined through equation (\ref{multipole_content}) for the vacuum case (where its value is $a_{\rm vac} = 1$ in the units listed in Table \ref{table_dimensions}), and calculate the relative difference as
	\beq
	\frac{\Delta a_1}{a_{\rm vac}} =
	\frac{a_1 - a_{\rm vac}}{a_{\rm vac}} \ ,
	\label{dipole_relative}
	\enq
shown as the dotted line in Figure \ref{fig_scaling}. The number of grid points (in each of the two dimensions) is shown in units of 25 (i.e.\ the total number of grid points is $N^2$). In all cases accuracy improves with increased number of grid points.

We have also applied our code to reproduce analytic solutions for the vacuum field and for the linear toroidal field ($T = sP$) for several multipoles. Such analytic solutions are discussed, for example, in Atanasiu et al.\ (2004) and Vigan\`{o} et al.\ (2011). In all cases the agreement is excellent, and typically around six significant digits. Thus, the linear solver is fairly robust.


	\begin{figure*}
	\centerline{\includegraphics[width=1\textwidth]{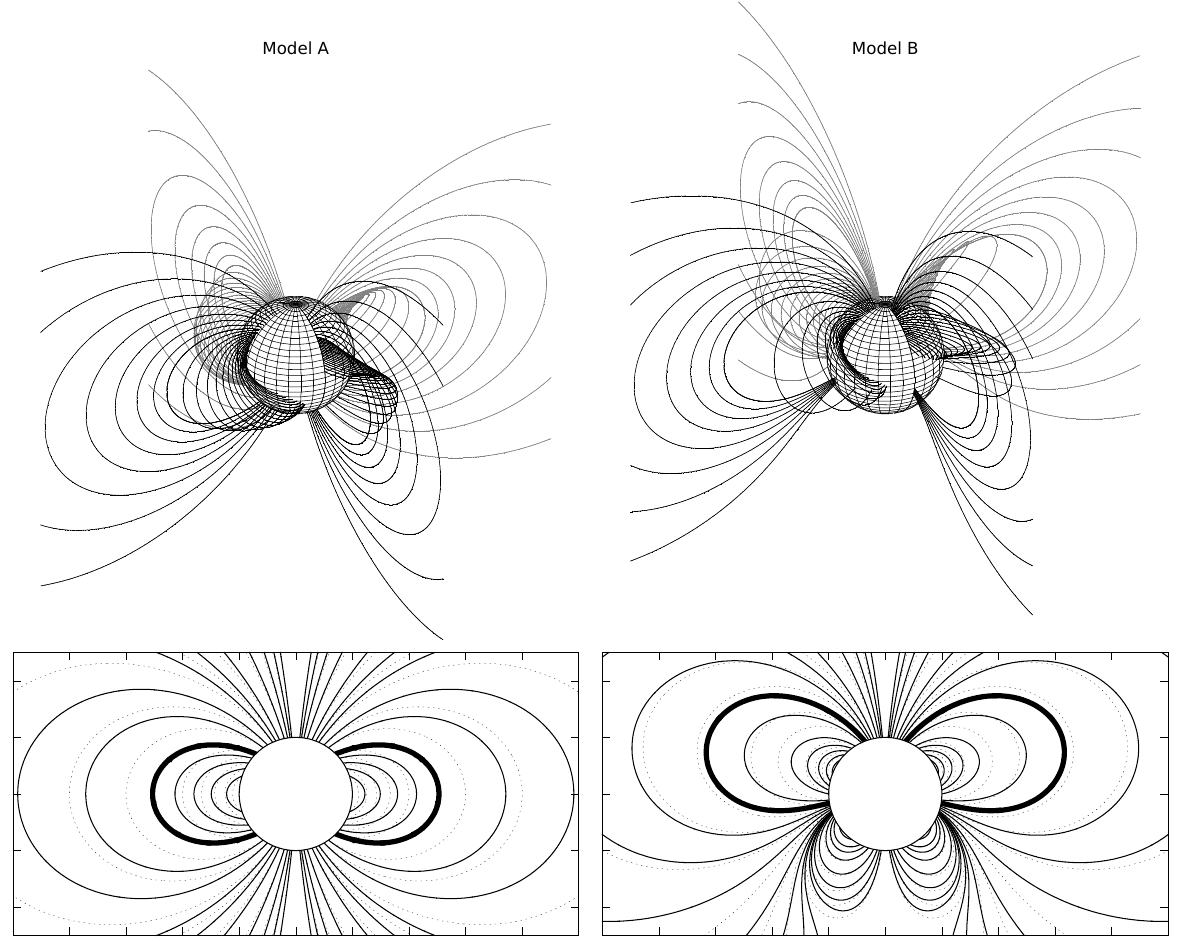}}
	\caption{{\bf Top}: field lines of two twisted magnetosphere models in three dimensions. The dipolar field (model A) is shown on the left, and the combination of a dipolar and quadrupolar field (model B) is shown on the right. The same set of field lines is reproduced in intervals of $\pi/2$ in the azimuthal angle $\phi$. In the current-free region (for $P < P_{\rm c}$) there is no twist and the field lines are coplanar. {\bf Bottom}: planar projection of some field lines in the top panels. The critical field line enclosing the toroidal region is highlighted with a thick line. The corresponding vacuum fields in both cases are shown as dotted lines in the background for comparison. The parameters and some calculated quantities for these two models are listed in Table \ref{table_parameters}.}
	\label{fig_3d}
	\end{figure*}

\section{Results and discussion}\label{section_results}
\subsection{Sample models for given $s$, $P_{\rm c}$ and $\sigma$}
We begin the discussion of our results by showing the magnetic field lines for two sample models in Figure \ref{fig_3d}. In the left panels the surface magnetic field is that of a dipole (hereafter referred to as \emph{model A}), while in the right panels it is a combination of a dipole and a quadrupole (labelled \emph{model B}). In both cases, the poloidal function at the surface can be expressed as a combination of the first two multipoles (cf.\ Appendix \ref{section_multipole}),
	\beq
	P(r,\theta) = (1-\mu^2) \left[ \frac{w P_1'(\mu)}{r}
	+ \frac{(1-w)P_2'(\mu)}{r^2} \right] \ .
	\label{P_combined}
	\enq
Here, the parameter $w$ controls the relative strength of the dipolar and quadrupolar components, and we take $w=1$ for the purely dipolar case and $w=0.5$ for the combined case. The toroidal field is of the form given by equation (\ref{T_of_P}), with $\sigma = 1$, and the complete list of parameters for the two models are listed in Table \ref{table_parameters}. We use a grid of $600\times 601$ points (the odd number for the angular grid is used in order to resolve the equator).

The magnetic energy of the current-free (vacuum) solution, with $P$ given as in equation (\ref{P_combined}) and $T=0$, is (cf.\ equation \ref{energy_vacuum2})
	\beq
	{E}_{\rm vac} = \frac{w^2}{3} + \frac{6(1-w)^2}{5} \ ,
	\enq
in the units listed in Table \ref{table_dimensions}. In particular, for a pure dipole ($w=1$) this gives $1/3$, and for $w=0.5$ it gives $23/60$. We can express the energies of the twisted magnetosphere models listed in Table \ref{table_parameters} relative to the energy of the vacuum solution. We thus obtain that models A and B contain $18\%$ and $13\%$ more energy than the corresponding vacuum solutions, respectively.

\begin{table}
\caption{List of parameters and numerical results for various quantities for the two models shown in Figure \ref{fig_3d}. The parameter $w$ is the weight defined in equation (\ref{P_combined}). The derived quantities are defined in \S\ref{section_auxiliary} and are expressed in the units listed in Table \ref{table_dimensions}. Numerical results are given to three significant digits.}
\label{table_parameters}
\center
\begin{tabular}{lcc}
\hline \hline
& model A & model B \\
\hline \hline
Parameters: & & \\
$w$ & 1 & 0.5 \\
$s$ & 1.6 & 1.15 \\
$P_{\rm c}$ & 0.5 & 0.25 \\
$\sigma$ & 1 & 1 \\
\hline\hline
Derived quantities: & & \\
Energy, ${E}$ & 0.393 & 0.432 \\
Energy, ${E}/{E}_{\rm vac}$ (\%) & 118\% & 113\% \\
Helicity, ${H}$ & 4.20 & 3.24 \\
Maximum twist, $\varphi_{\rm max}$ & 1.22 & 1.07 \\
Dipole strength, $a_1$ & 1.22 & 0.699 \\
Quadrupole strength, $a_2$ & - & 0.655 \\
\hline \hline
\end{tabular}
\end{table}

	\begin{figure*}
	\centerline{\includegraphics[width=1\textwidth]{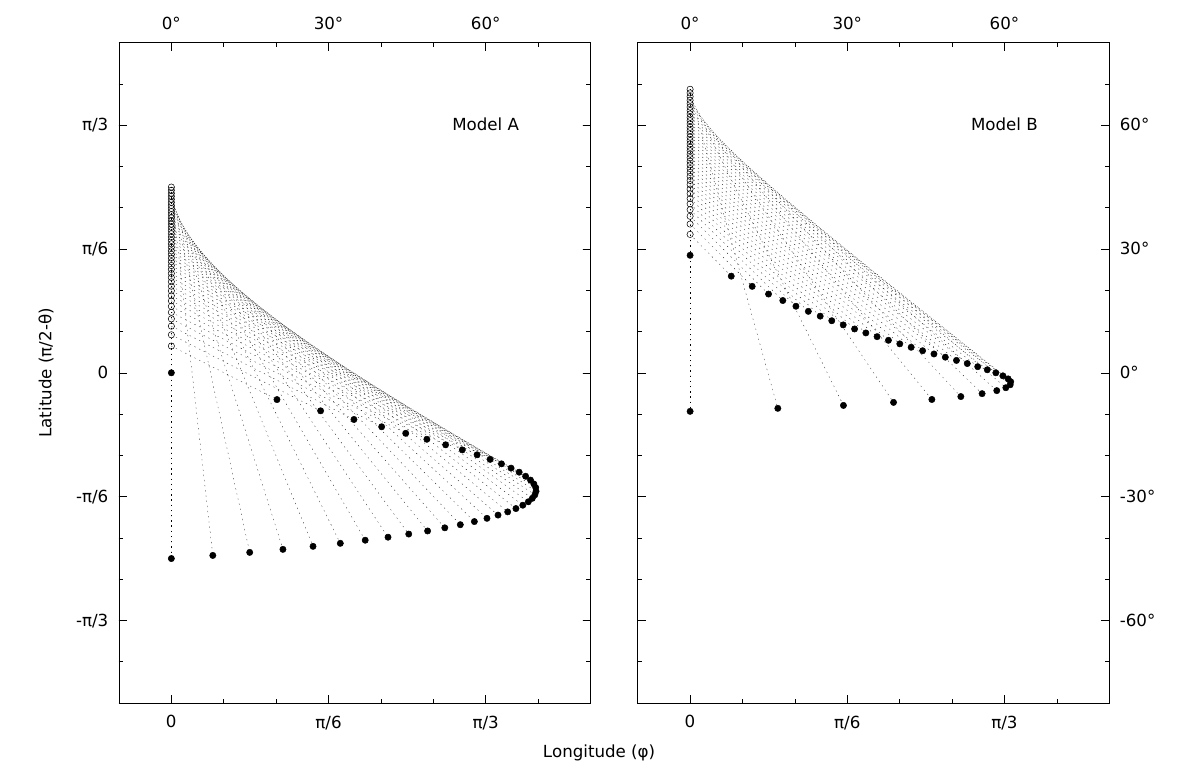}}
	\caption{Surface map of the footprints of field lines on the stellar surface, in terms of latitude ($\pi/2-\theta$, vertical) and longitude (or, azimuthal angle, which in this case is the same as the twist $\varphi$, horizontal), in radians and degrees, for the same models as in Figure \ref{fig_3d}. Only field lines in the toroidal region are shown. (Purely poloidal fields have zero twist.) Each field line has two footprints: the points where the field lines come out of the surface are shown with empty circles, and the points where the field lines reenter the surface are shown with full circles. The points of exit and reentry for the same line are connected with a dotted line, which is also the projection of the three dimensional field line onto the surface. Note that at the boundary of the toroidal region and at the central point where the field line length is zero (which corresponds to the equator for the dipolar case in model A, and to $\approx 28.5^\circ$ for model B) the twist goes to zero. The maximum twist for model A is $\varphi_{\rm max} \approx 1.22$ rad, and for model B it is $\varphi_{\rm max} \approx 1.07$ rad (Table \ref{table_parameters}).}
	\label{fig_twist}
	\end{figure*}

	\begin{figure}
	\includegraphics[width=0.5\textwidth]{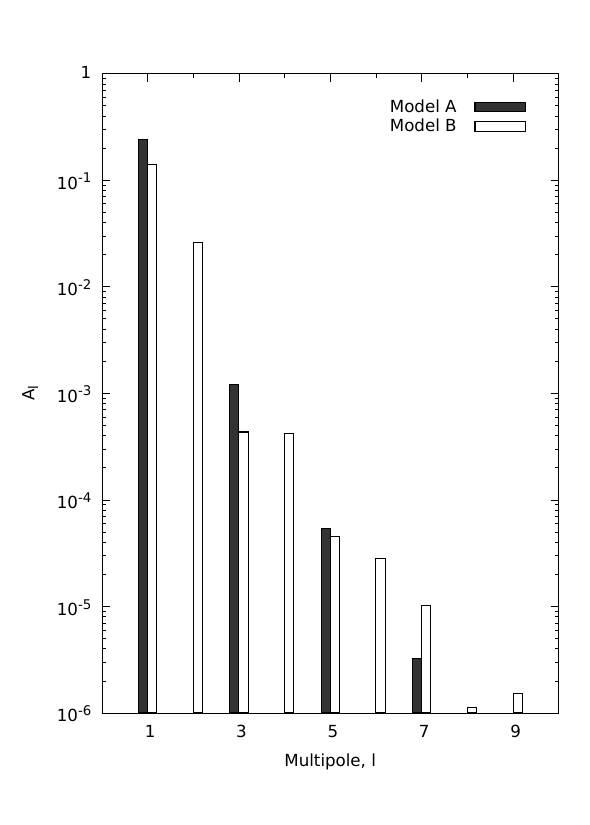}
	\caption{Multipole content of the vacuum field for the two models shown in Figure \ref{fig_3d}. In both cases, $A_l$ are the amplitudes from the multipole expansion (equation \ref{multipole}) at the external radius $r_{\rm out} = 5$, where the field is current-free. Note that the even multipoles are absent (except for numerical noise) for model A, and the amplitudes of higher multipoles decrease rapidly with $l$ for both models.}
	\label{fig_multipole}
	\end{figure}

The {\it twist} is defined as the azimuthal extent of a field line through equation (\ref{twist}). Figure \ref{fig_twist} shows a projection of field lines on the stellar surface, and illustrates how twist depends on latitude. Outside the toroidal region, the twist is zero as there is no toroidal field. The twist increases linearly with the toroidal field strength (which increases towards the middle of the toroidal region), but it also depends (non-linearly) on the field line length, which becomes vanishingly small in the same limit. Therefore, the maximum twist is reached for an intermediate angle, and then drops back to zero as we approach the equator (for model A) or $\approx 28.5^\circ$ (for model B), corresponding to the maximum of the function $P$ defined in equation (\ref{P_combined}). Interestingly, for both models, this maximum twist is similar (1.22 and 1.07 rad, respectively; see Table \ref{table_parameters}). In the following sections we extend the discussion about this point.

The last two quantities listed in Table \ref{table_parameters} are the dipole and quadrupole content of the models. For model A, the dipole component at the surface is 1, but the currents in the twisted magnetosphere augment this to $a_1 = 1.22$. Similarly, for model B the surface dipole and quadrupole components of $0.5$ each are augmented to $a_1 = 0.699$ and $a_2 = 0.655$, respectively. In Figure \ref{fig_multipole} we plot the coefficients $A_l$ of the multipole expansion at $r_{\rm out} = 5$, showing how higher multipoles drop off quickly  (as $r^{-l}$).  We also note that the symmetry of the surface field is preserved, and the magnetospheric currents in model A only generate odd multipolar components. Note that the multipole coefficients rescaled to their surface values ($a_l$ defined in equation \ref{multipole_content}) are independent of the radius where the expansion is carried out, as long as it lies beyond the region containing the currents.

\subsection{Dependence of the results on the parameters $s$, $P_{\rm c}$ and $\sigma$}\label{section_one_parameter}
In the dipolar model A, the parameter $s = 1.6$ was chosen to be very close to the maximum value for which convergence could be reached. For values $s \gtrsim 1.62$,  we could not find any solutions. In this subsection, we explore how this maximum value of $s$ is correlated with the other parameters $P_{\rm c}$ and $\sigma$, and how the physical quantities (energy, helicity, twist and dipole content) depend on these parameters. In all models considered in this section, we impose a dipolar field for the poloidal function at the stellar surface.

\begin{figure*}
	\centerline{\includegraphics[width=0.95\textwidth]{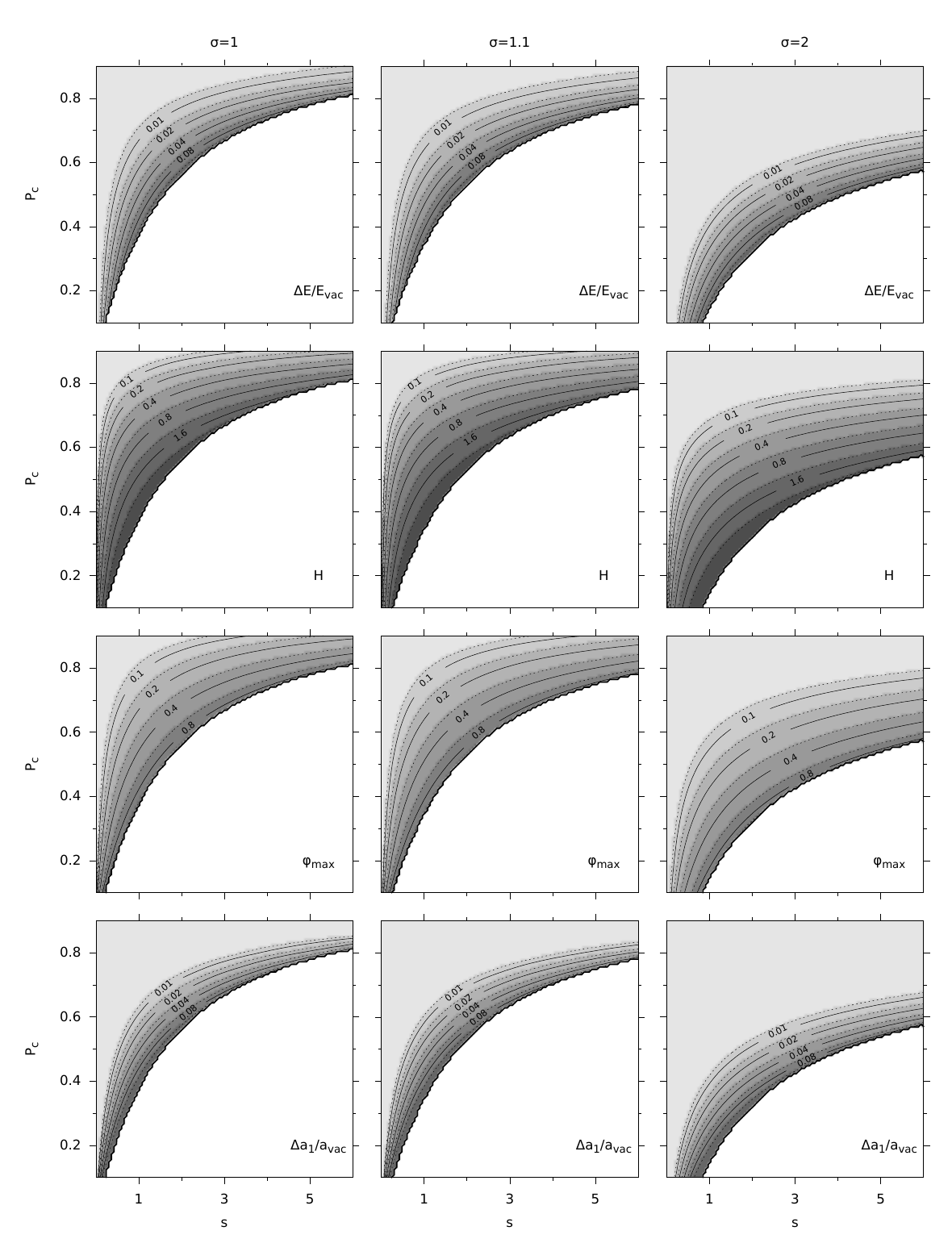}}
	\caption{Relative energy, helicity, maximum twist and relative dipole strength for three models with $\sigma = 1$, $1.1$ and $2$ as functions of the parameters $s$ and $P_{\rm c}$. The regions where solutions have not been found are shown in white. The levels of the contours are indicated in the plots. Note that, in particular, the contours of the maximum twist seem to align remarkably well with the boundary where convergence fails.}
	\label{fig_map}
	\end{figure*}

In Figure \ref{fig_map}, we show contours of relative energy, helicity, maximum twist and relative dipole strength, in a two dimensional parameter space (as functions of $s$ and $P_{\rm c}$), and for three models with $\sigma=1$ (left panels), $1.1$ (central panels), and $2$ (right panels). The relative energy and dipole strength are calculated with respect to the vacuum solution, through equations (\ref{energy_relative}) and (\ref{dipole_relative}), respectively. Note that both of these quantities represent an increase with respect to the vacuum case. The plots are produced for grids of around $200\times 200$ points in radius and angle, where the error in the numerical calculation of the energy for the vacuum case ($s=0$) is of the order of $0.1\%$ (cf.\ Figure \ref{fig_scaling}).

At first sight, we can detect a few interesting features. First, the energy and dipole strength of models with twisted magnetospheres are increased by moderate amounts, typically in the vicinity of $10\%$, with respect to their vacuum values, with the largest increases that we have been able to find being around $25\%$ for the energy, and up to $40\%$ for the dipole strength. The helicity of these models is found to reach values of up to $\sim 5$, while the maximum twist is typically $\lesssim 1.5$.

The most intriguing fact is that, irrespective of the large variations in the parameter space (in $s$, $P_{\rm c}$ and $\sigma$), all models seem to fail to find new solutions when the maximum twist is around $1.2 - 1.5$. We note that the white region on the lower right part of each plot (corresponding to the parameter space where convergence fails) is remarkably well aligned with some quantities, but especially with the maximum twist. Apparently, very different models, whether involving large volumes (small $P_{\rm c}$), but limited to small $s$, or involving small volumes ($P_{\rm c}$ close to 1), but allowing for large $s$, are limited by the same reason: when the maximum twist of any field line reaches a critical value of $\approx 1.2 - 1.5$ no more solutions are found.

The plots for $\sigma=1.1$ demonstrate the close resemblance to the case for $\sigma=1$. Increasing $\sigma$ further concentrates the toroidal field near the equator, and consequently diminishes its effect on the structure of the poloidal field. Therefore, solutions span a larger area of the parameter space in $s$ and $P_{\rm c}$, as can be seen in the plots for $\sigma=2$, yet, crucially, the above conclusion for the maximum twist still holds.

\begin{figure}
	\centerline{\includegraphics[width=0.5\textwidth]{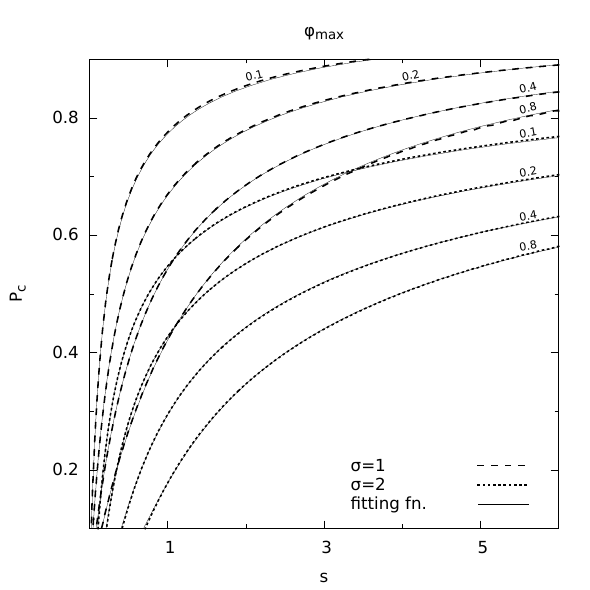}}
	\caption{Fitting functions for the contours of the maximum twist. The same contours for the maximum twist $\varphi_{\rm max}$ are shown for $\sigma=1$ (dashed lines) and $\sigma=2$ (dotted lines), as in Figure \ref{fig_map}. The fitting functions are shown with solid lines, and are of the form given by equation (\ref{fit}). The parameters for each line are listed in Table \ref{table_fit}.}
	\label{fig_mapfit}
	\end{figure}

We also find that the contours shown in Figure \ref{fig_map} are fairly well described by a function of the form
	\beq
	s = \frac{\gamma P_{\rm c}^m}{(1-P_{\rm c})^n} \ .
	\label{fit}
	\enq
The three unknown parameters $\gamma$, $m$ and $n$ can be determined through a best fit, or more simply, by imposing three equations at three arbitrary points along a given contour. Following the latter procedure, we find that the energy, helicity and dipole strength contours can be quite well represented by such a function, where the parameters need to be determined individually for each line. Even more spectacular is the fit to the maximum twist. The contour lines and some sample fits are shown for this case in Figure \ref{fig_mapfit}, and the parameters of these fits are listed in Table \ref{table_fit}. In general, the parameters are also functions of $\varphi_{\rm max}$. From an inspection of the values listed in the table, we observe that $\gamma$ is approximately linear with $\varphi_{\rm max}$, while $m$ and $n$ do not vary much.

We next discuss in more detail the dependence of the solutions on each of the two parameters $s$ and $P_{\rm c}$.

\begin{table}
\caption{List of parameters for the fitting function given by equation (\ref{fit}) for the contours of the maximum twist $\varphi_{\rm max}$ for $\sigma = 1$ and $\sigma = 2$ shown in Figure \ref{fig_mapfit}.}
\label{table_fit}
\center
\begin{tabular}{ccccccccc}
\hline \hline
& \ & \multicolumn{3}{c}{$\sigma = 1$} & \ & \multicolumn{3}{c}{$\sigma = 2$} \\
\hline \hline
$\varphi_{\rm max}$ & & $\gamma$ & $m$ & $n$ & & $\gamma$ & $m$ & $n$ \\
\hline \hline
0.1 & & 0.155 & 0.892 & 1.41 & & 0.156 & 0.284 & 2.56 \\
0.2 & & 0.315 & 0.903 & 1.38 & & 0.310 & 0.280 & 2.53 \\
0.4 & & 0.661 & 0.937 & 1.26 & & 0.621 & 0.296 & 2.41 \\
0.8 & & 1.35  & 0.983 & 1.00 & & 1.13  & 0.309 & 2.12 \\
\hline\hline
\end{tabular}
\end{table}

	\begin{figure*}
	\centerline{\includegraphics[width=1\textwidth]{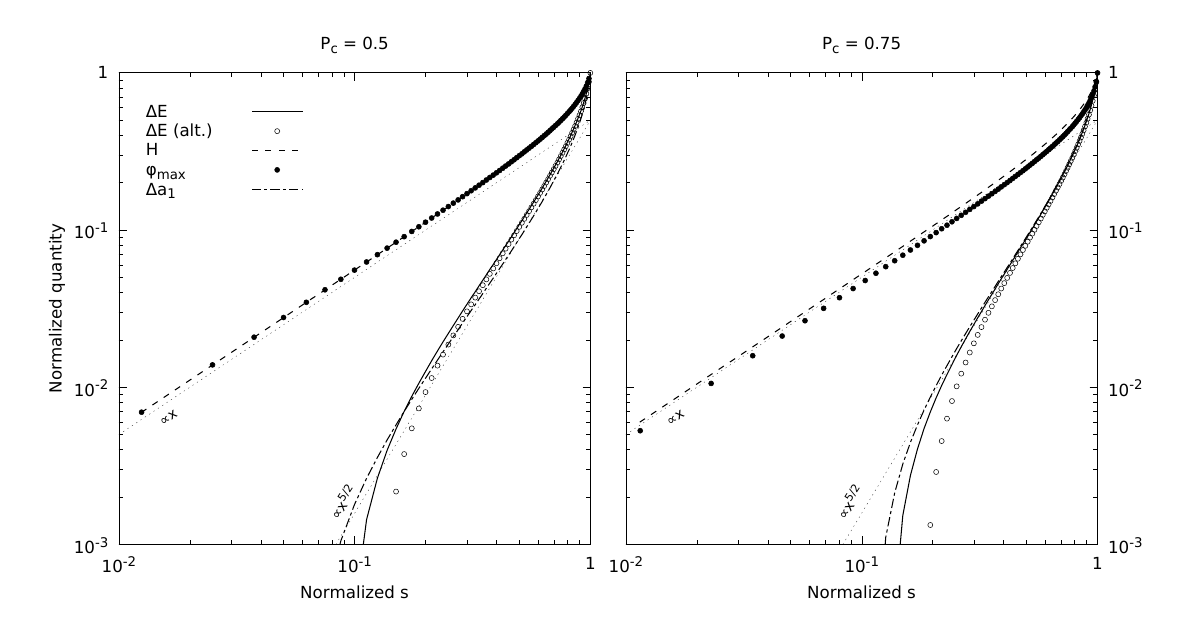}}
	\caption{Log--log plots of normalized energy, helicity, maximum twist and dipole strength as functions of normalized $s$ for $P_{\rm c} = 0.5$ (left) and $P_{\rm c} = 0.75$ (right). In both cases $\sigma = 1$. The energy and dipole strength are expressed relative to their vacuum values as defined through equations (\ref{energy_relative}) and (\ref{dipole_relative}), respectively. All quantities have been rescaled by their largest values, listed in Table \ref{table_largest}. For reference, we have included trend lines of linear and $x^{5/2}$ dependence. Note that all quantities approach their vacuum values (in this case, 0) as $s \to 0$.}
	\label{fig_parameter_s}
	\end{figure*}

	\begin{figure*}
	\centerline{\includegraphics[width=1\textwidth]{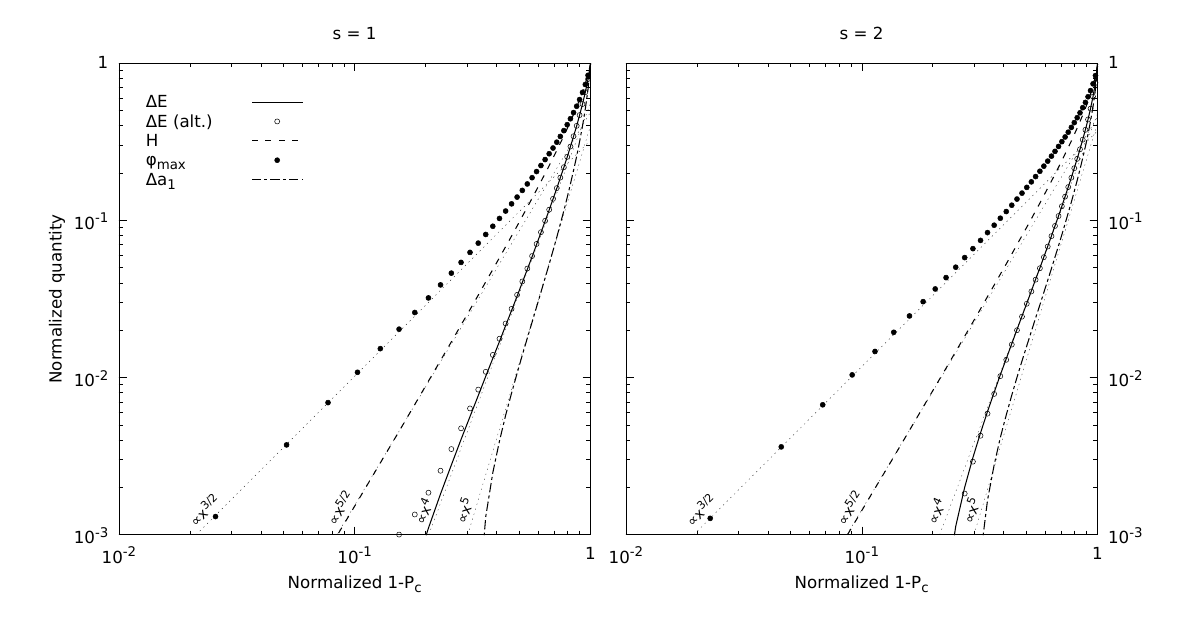}}
	\caption{Log--log plots of normalized energy, helicity, maximum twist and dipole strength as functions of normalized $1 - P_{\rm c}$ for $s=1$ (left) and $s=2$ (right). In both cases $\sigma = 1$. As in Figure \ref{fig_parameter_s}, we plot the relative energy and dipole strength and normalize all quantities by their largest values (which for $P_{\rm c}$ corresponds to its smallest value), listed in Table \ref{table_largest}. The vacuum case is retrieved in the limit $P_{\rm c} \to 1$.}
	\label{fig_parameter_pc}
	\end{figure*}

\subsubsection{Dependence on $s$}
In Figure \ref{fig_parameter_s} we show the relative energy, helicity, maximum twist and relative dipole strength as a function of $s$ for a fixed $P_{\rm c}$. Two cases for $P_{\rm c} = 0.5$ and $P_{\rm c} = 0.75$ are shown and in both cases we set $\sigma = 1$. We plot all quantities normalized to their maximum values (which are always reached just as convergence fails, and are listed in Table \ref{table_largest}). For comparison, we show the relative energy increase calculated with the two methods described in \S\ref{section_energy2}: as a volume integral of the magnetospheric region plus a surface integral for the region external to the outer numerical boundary (shown with a solid line), or, alternatively, as a surface integral (shown with empty circles). Both are in good agreement, save for numerical errors due to the finite resolution.

Clearly, the (normalized) helicity $H$ and maximum twist $\varphi_{\rm max}$ are closely correlated. In the limit when the poloidal field lines are not strongly modified by the presence of magnetospheric currents (for small $s$), both quantities grow linearly with $s$, as should be expected since both depend linearly on the toroidal field $B_\phi$. The deviation from the linear dependence of $H$ and $\varphi_{\rm max}$ is visible only near the maximum value of $s$. On the other hand, the relative energy $\Delta E$ and the relative dipole strength $\Delta a_1$ both scale as $s^{5/2}$. This is an indication that, to leading order, the increase in the energy of the magnetosphere model is mostly due to the amplification of the dipolar field. If the energy increase could be attributed to the toroidal field, it should scale as $\Delta E \propto B_\phi^2 \propto s^2$. Conversely, if the energy increase is due to the increase of the dipolar field strength, we would expect to have $\Delta E \propto \Delta (B_{\rm pol}^2) \propto (a_1+\Delta a_1)^2 - a_1^2 \propto \Delta a_1$, since we are still in the regime $\Delta a_1 \ll a_1$. This explains why both normalized functions ($\Delta E$ and $ \Delta a_1$) scale in the same way.

\subsubsection{Dependence on $P_{\rm c}$}
In Figure \ref{fig_parameter_pc}, we show the dependence of the same quantities on $1-P_{\rm c}$, for $s=1$ (left) and $s=2$ (right). As in Figure \ref{fig_parameter_s}, we normalize all quantities to their maximum values (listed in Table \ref{table_largest}) attained at the critical value of $P_{\rm c}$ beyond which the numerical solution does not converge.

For a dipole field, $P_{\rm c}$ has a maximum value of 1 at the equator on the stellar surface. In this case, the toroidal field is confined to a single point and the field configuration reduces to the vacuum case (as is evident in the figure for the limit $P_{\rm c} \to 1$). As $P_{\rm c}$ is decreased from 1 (i.e.\ as $1-P_{\rm c}$ is increased from 0), the volume occupied by the toroidal field increases and subsequently the poloidal field lines are increasingly modified with respect to the vacuum configuration. Beyond some minimum $P_{\rm c}$ (listed in Table \ref{table_largest}) no solutions are found.

Plotting the quantities as functions of $1-P_{\rm c}$ (rescaled by its largest value) reveals some nice scalings. In particular, the maximum twist scales as $x^{3/2}$, the total helicity as $x^{5/2}$, the relative energy (with respect to the vacuum) as $x^{4}$, and the relative dipole strength as $x^{5}$. When $P_{\rm c}$ is near 1, the field lines are very close to the equator and very short, and consequently higher resolution is required in order to maintain accuracy. This explains the observed divergence in the scalings for small values of $1-P_{\rm c}$. 

This difference in the scalings of the maximum twist and helicity can be attributed to the larger volume occupied by the magnetosphere as $1-P_{\rm c}$ increases. The maximum twist depends only on the toroidal field strength and the field line length, but the helicity is a volume integrated quantity. In Appendix \ref{section_construct} we present a mathematical construct which allows us to analytically calculate the helicity and maximum twist for a simple model. In addition to being a useful tool for performing numerical checks, this model is also valid in the limit of weak toroidal fields where the poloidal field structure remains nearly unchanged. Taking the corresponding limit for small $1-P_{\rm c}$, we indeed verify that the scalings of helicity and maximum twist shown in Figure \ref{fig_parameter_pc} are correct (cf.\ equation \ref{limiting_values}).

\begin{table}
\caption{List of the values of the quantities used in the normalization of Figures \ref{fig_parameter_s} and \ref{fig_parameter_pc} (in the units listed in Table \ref{table_dimensions} and expressed to three significant digits). The vacuum values of the energy and dipole strength used to calculate the relative values defined through equations (\ref{energy_relative}) and (\ref{dipole_relative}) are ${E}_{\rm vac} = 1/3$ and $a_{\rm vac} = 1$, respectively. The numbers shown in parentheses are kept fixed in their respective columns.}
\label{table_largest}
\center
\begin{tabular}{lcccccc}
\hline \hline
& \ & \multicolumn{2}{c}{Figure \ref{fig_parameter_s}} & \ & \multicolumn{2}{c}{Figure \ref{fig_parameter_pc}} \\
\hline \hline
Quantity & & Left & Right & & Left & Right \\
\hline \hline
$s$	& & 1.62 & 4.35 & & (1) & (2) \\
$P_{\rm c}$	& & (0.5) & (0.75) & & 0.376 & 0.560 \\
${E}$ & & 0.400 & 0.374 & & 0.380 & 0.396 \\
$H$ 	& & 4.51 & 1.96 & & 4.41 & 3.90 \\
$\varphi_{\rm max}$ & & 1.31 & 1.11 & & 1.21 & 1.22 \\
$a_1$ & & 1.25 & 1.08 & & 1.25 & 1.20 \\
\hline\hline
\end{tabular}
\end{table}

\section{Conclusions}\label{section_conclusions}
In this work, we study the properties of force-free magnetosphere models, which satisfy appropriate boundary conditions at the stellar surface, at the axis, and at infinity. In particular, we impose that the magnetic field match smoothly to a current-free (vacuum) solution at some large external radius. Our models are designed in a way that ensures there are no pathological surface currents at any of the interfaces. The boundary condition at the stellar surface allows us to prescribe any poloidal function $P$ and toroidal function $T(P)$, where, for the latter, we assume the form given by equation (\ref{T_of_P}). The sample solutions shown in Figure \ref{fig_3d} correspond to dipolar and mixed (dipolar plus quadrupolar) configurations. Clearly, these models are very specific, but they serve as an illustration of our method. In general, changing the surface profiles of $P$ and $T$ would affect the form of the resulting magnetosphere.

We have carried out an extensive parametric study revealing how important quantities (energy, helicity, twist, and multipole content) vary with the different parameters describing our model. We find that the total dipolar moment can be increased by up to $40\%$ with respect to a vacuum model with the same surface magnetic field. This is simply reflecting the contribution of the magnetospheric currents to the global magnetic field. Thus, the estimates of the surface magnetic field based on properties of the large-scale dipole (e.g.\ braking torque) are slightly overestimating the surface value. We also find a moderate increase in the total energy of the model with respect to the vacuum solution of up to $25\%$. We attribute most of this energy increase to the higher dipole moment, rather than to the energy stored in the toroidal field, since the volume occupied by the toroidal field is not large and the volume integrated poloidal component (which extends to very long distances) always dominates.

We have also found the interesting result of the existence of a critical twist ($\varphi_{\rm max} \approx 1.2-1.5$ rad, for the models studied). This idea has been suggested by different authors in other contexts and with other approaches. For example, by performing numerical simulations of resistive MHD applied to the disruption of coronal arcades, Mikic \& Linker (1994) arrived at the result that there are no more force-free equilibria beyond a maximum twist (maximum \emph{shear} in their terminology) of $\varphi_{\rm max} \approx 1.6$ rad, for their particular functional form of the applied shear. This fact has interesting implications: if some internal mechanism (for example, MHD instabilities, Hall drift, or ambipolar diffusion) results in a slow transfer of helicity to the magnetosphere (thus increasing the twist), there appears to be a critical value of the twist beyond which force-free solutions no longer exist. Further increasing this value might result in the sudden disruption of the magnetospheric loops and may be at the origin of phenomena such as soft gamma repeaters (SGRs) or X-ray bursts.

In general, our results agree qualitatively with previous works (Fujisawa \& Kisaka 2014; Glampedakis et al.\ 2014; Pili et al.\ 2015), with some minor differences. Unlike Fujisawa \& Kisaka (2014) and Pili et al.\ (2015), we do not find solutions with disconnected toroidal loops, which, as is argued in both papers, are probably unstable. The formation of such loops seems to be a consequence of the fact that they make use of a different iteration scheme in their work, where they fix the size of the toroidal region while allowing $P_{\rm c}$ to vary between iterations. When such disconnected loops are formed, in principle, it is possible to inject more helicity and twist into the magnetosphere, thus also increasing the dipole content and, in particular, the energy budget available for fast, global magnetospheric activity. However, such solutions are not found when carrying out iterations for a fixed value of $P_{\rm c}$, while allowing the size of the toroidal region to vary, as in our case. Instead, our solutions always seem to converge to magnetospheres with smaller toroidal regions, and with field lines connected to the stellar surface. Thus, it is plausible that the disconnected loops represent highly localized regions in the parameter space where the Grad--Shafranov equation has more than one acceptable solution, with those presented in this paper representing the lower energy (and helicity and twist) solutions, and thus being energetically favorable over the seemingly unstable disconnected ones. At the moment this remains purely as a conjecture, and more work needs to be carried out in order to better understand the possible degeneracy of the solutions and its implications.

In this work, we do not solve for the internal magnetic field of the star, and instead impose boundary conditions on the poloidal and toroidal functions at the stellar surface. While solutions of the Grad--Shafranov equation are useful for constructing equilibria in the interior of barotropic stars, these equilibria are unlikely to be stable (Markey \& Tayler 1973; Tayler 1973; Lander \& Jones 2012; Mitchell et al.\ 2015). A solution to the stability problem may be the so-called non-barotropic fluids where stable stratification due to composition and entropy gradients can balance some of the instabilities (Reisenegger 2009; Akg\"{u}n, et al.\ 2013). In this case, solutions of the Grad--Shafranov equation are no longer required, and there is a wider range of acceptable magnetic field configurations in equilibrium. Either way, once the long-term evolution of the internal magnetic field due to the Hall, Ohmic and ambipolar terms kicks in, it is clear that no matter what the initial field is chosen to be, the magnetic field at subsequent snapshots will not be a solution of the Grad--Shafranov equation. Thus, from the perspective of long-term evolution, the interior field is determined through the evolution equations, while the magnetosphere relaxes on much shorter timescales to a force-free configuration matching the surface boundary conditions. This paper is the preliminary step of further future work, where we will combine this family of magnetosphere models with the long-term magnetic field evolution inside the star obtained from the code described  by Vigan\`{o}, Pons \& Miralles (2012), to explore the influence of the magnetosphere on the magneto-thermal evolution of neutron stars.

\section*{Acknowledgements}
This work is supported in part by the Spanish MINECO grants AYA2013-40979-P, AYA2013-42184-P, and AYA2015-66899-C2-2-P, the grant of Generalitat Valenciana PROMETEOII-2014-069, the European Union ERC Starting Grant 259276-CAMAP, and by the New Compstar COST action MP1304.



\appendix
\section{Energy of force-free fields}\label{section_proof}
The energy of any general force-free magnetic field can be expressed entirely in terms of surface integrals. Following Chandrasekhar (1981), consider the integral of the work done by the Lorentz force,
	\beq
	\begin{split}
	& \int\m{r}\cdot(\m{\nabla}\times\m{B})\times\m{B} \, dV =
	\int \epsilon_{ijk}\epsilon_{jlm} r_i B_k (\nabla_l B_m) dV \\
	= & \ \epsilon_{ijk}\epsilon_{jlm} \left[\oint r_i B_k B_m dS_l
	- \int \nabla_l(r_i B_k) B_m dV\right] \\
	= & - \oint r_i B_k^2 dS_i + \oint r_i B_i B_k dS_k \\
	& - \int \nabla_k(r_i B_k) B_i dV + \int \nabla_i(r_i B_k) B_k dV \\
	= & - \oint r_i B_k^2 dS_i + \oint r_i B_i B_k dS_k \\
	& + 2 \int B_k^2 dV + \frac{1}{2} \int r_i \nabla_i B_k^2 dV \\
	= & - \frac{1}{2} \oint r_i B_k^2 dS_i + \oint r_i B_i B_k dS_k
	+ \frac{1}{2} \int B_i^2 dV \ .
	\end{split}
	\enq
Here, we carry out a second integration by parts in the last line and make use of the relation between the Levi--Civita symbol $\epsilon_{ijk}$ and the Kronecker delta $\delta_{ij}$,
	\beq
	\epsilon_{ijk} \epsilon_{klm} = \delta_{il} \delta_{jm} - \delta_{im} \delta_{jl} \ ,
	\enq
as well as the derivatives of the radial vector,
	\beq
	\nabla_i r_j = \delta_{ij} \ ,
	\enq
which, in particular, for the divergence gives $\m{\nabla}\cdot\m{r} = \nabla_i r_i = \delta_{ii} = 3$. In vector notation, the final result can be expressed as
	\beq
	\begin{split}
	& \int \m{r}\cdot(\m{\nabla}\times\m{B})\times\m{B} \, dV = \\
	& = - \frac{1}{2} \oint B^2 (\m{r}\cdot d\m{S})
	+ \oint (\m{r}\cdot \m{B}) (\m{B}\cdot d\m{S})
	+ \frac{1}{2} \int B^2 \, dV \ .
	\end{split}
	\label{chandra}
	\enq
For a force-free field the left-hand side vanishes and we can write the magnetic energy purely in terms of surface terms,
	\beq
	8\pi {E} = \oint B^2 (\m{r}\cdot d\m{S})
	- 2 \oint (\m{r}\cdot \m{B}) (\m{B}\cdot d\m{S}) \ ,
	\label{energy_ff}
	\enq
which is reproduced as equation (\ref{energy}) in this text.

Note that in order to be able to use this formula, the magnetic field should be at least once differentiable. This, in turn, implies that the poloidal function $P$ should be at least twice differentiable and the toroidal function $T$ should be at least once differentiable (cf.\ equation \ref{mag_PT}). Thus, for example, when surface currents are present and the magnetic field is not continuous, this formula cannot be applied.

\section{Energy of current-free (vacuum) fields}\label{section_vacuum}
Current-free fields satisfy $\m\nabla\times\m{B} = 0$, implying that the magnetic field can be written in terms of some scalar potential through $\m{B} = \m\nabla\Psi$. Since the magnetic field is divergenceless (solenoidal), the function $\Psi$ is given as a solution of Laplace's equation, $\nabla^2 \Psi = 0$. It can also be reconstructed directly from the multipolar expansion of the poloidal function $P$. Using  equation (\ref{multipole}) and that $\m{B} = \m\nabla\Psi = \m\nabla P\times\m\nabla\phi$, we get
	\beq
	\Psi(r,\theta) = \sum_{l=1}^\infty A_l'(r) P_l (\mu) \ ,
	\label{psi}
	\enq
where the radial functions $A_l(r)$ are given by equation (\ref{vacuum_coeff}) for the vacuum case.

Using Gauss's (divergence) theorem, the magnetic energy can then be written as (Marchant, Reisenegger \& Akg\"{u}n 2011)
	\beq
	\begin{split}
	8\pi {E} & = \int (\m\nabla\Psi)^2 \, dV \\
	& = \int \m\nabla\cdot(\Psi\m\nabla\Psi) \, dV
	= \oint \Psi\m\nabla\Psi\cdot d\m{S} \ .
	\label{energy_vacuum}
	\end{split}
	\enq
Thus, we can write the energy of a current-free field either as in equation (\ref{energy_ff}) or as in here.
Note that the vectors $\m\nabla P$, $\m\nabla\phi$ and $\m{B} = \m\nabla\Psi = \m\nabla P\times\m\nabla\phi$ are mutually orthogonal, and subsequently, depending on the shape of the surface over which the integration is carried out, using one or the other formula may be more advantageous.

For reference, the energies stored in the vacuum dipole and quadrupole are
	\beq
	{E}_{\rm dip} = \frac{B_{\rm o}^2 R_\star^3}{3} \mtext{and}
	{E}_{\rm quad} = \frac{6 B_{\rm o}^2 R_\star^3}{5} \ ,
	\label{energy_vacuum2}
	\enq
respectively.

\section{Multipole expansion}\label{section_multipole}
The poloidal function $P$ can be expanded in multipoles as
	\beq
	P(r,\theta) = (1-\mu^2)\sum_{l=1}^\infty A_l(r) P_l'(\mu) \ .
	\label{multipole}
	\enq
Here $P_l$ are the Legendre polynomials, which are solutions of the Legendre differential equation
	\beq
	[(1-\mu^2)P_l'(\mu)]' = - l(l+1)P_l(\mu) \ .
	\label{Legendre_diff}
	\enq
The dipole corresponds to $l=1$ and the quadrupole to $l=2$. The corresponding Legendre polynomials are
	\beq
	P_1(\mu) = \mu \mtext{and} P_2(\mu) = \frac{3\mu^2 - 1}{2} \ .
	\enq

In general, the radial functions $A_l(r)$ are determined from the Grad--Shafranov equation (equation \ref{GS_eq}). For the current-free (vacuum) case ($\triangle_{\rm GS} P = 0$), using the definition of the Grad--Shafranov operator (equation \ref{GS_op}) and the Legendre differential equation (equation \ref{Legendre_diff}), we get
	\beq
	(1-\mu^2) \sum_{l=1}^\infty \left[ A_l''(r) 
	- \frac{l(l+1)}{r^2} A_l(r) \right] P_l'(\mu) = 0 \ .
	\enq
In this case the multipoles are completely decoupled (which is not necessarily the case in general) and the solutions are of the form
	\beq
	A_l(r) = a_l r^{-l} + b_l r^{l+1} \ .
	\label{vacuum_coeff}
	\enq
In the stellar exterior we need to take $b_l = 0$. Thus, the vacuum dipole and quadrupole fields in the exterior are of the form
	\beq
	\begin{split}
	P_{\rm dip} & = a_1 \frac{\sin^2\theta}{r} \ , \\
	P_{\rm quad} & = a_2 \frac{3\sin^2\theta\cos\theta}{r^2} \ .
	\end{split}
	\label{dip_quad}
	\enq

\section{Mathematical construct}\label{section_construct}
There is no general analytic solution for the force-free case with both poloidal and toroidal components of the form considered in this text. Nevertheless, it is possible to construct a mathematical model which although not realistic can still serve for performing numerical checks and as a useful approximation in some limiting cases.

Assume that the poloidal field is that of a vacuum dipole, which in the units listed in Table \ref{table_dimensions} can be expressed as (equation \ref{dip_quad})
	\beq
	P(r,\theta) = \frac{\sin^2\theta}{r} \ ,
	\enq
while the toroidal field is still given through equation (\ref{T_of_P}) for some values of the three parameters $s$, $P_{\rm c}$ and $\sigma$. We choose $\sigma=1$ which is the easiest to calculate analytically. This solution corresponds to the limit of the weak toroidal field, and can be used as an indication of how various quantities that depend on the toroidal field behave in that limit.

The volume occupied by the toroidal field (in the magnetosphere) is a function of the critical field line $P_{\rm c}$ (which also defines the integration boundary) and in this case is given through
	\beq
	\begin{split}
	V_{\rm tor} = & \frac{4\pi \sqrt{1 - P_{\rm c}}}{3P_{\rm c}^3} \\
	& \times \left[P_{\rm c}(1-P_{\rm c}^2)
	+ \frac{3}{5}(1-P_{\rm c})^2 - \frac{1}{7}(1-P_{\rm c})^3\right] \ .
	\end{split}
	\enq
Similarly, helicity as defined through equation (\ref{helicity}) can be determined as a function of $s$ and $P_{\rm c}$,
	\beq
	\begin{split}
	H = & \frac{16\pi s P_{\rm c}}{3} \\
	& \times \left[\frac{1+2P_{\rm c}}{P_{\rm c}}\sqrt{1-P_{\rm c}}
	- 3\ln (1+\sqrt{1-P_{\rm c}}) + \frac{3}{2} \ln P_{\rm c} \right] \ .
	\end{split}
	\enq
The twist defined through equation (\ref{twist}) for a field line $P_{\rm o}$ which lies within the toroidal region (i.e.\ for $P_{\rm c} \leqslant P_{\rm o} \leqslant 1$) is a function of $s$ and $P_{\rm c}$, as well as $P_{\rm o}$
	\beq
	\varphi = \frac{2s(P_{\rm o}-P_{\rm c})\sqrt{1-P_{\rm o}}}{P_{\rm o}^2} \ .
	\enq
The value of $P_{\rm o}$ for which the twist becomes a maximum ($\varphi_{\rm max}$) is given through
	\beq
	P_{\rm o} = \frac{3P_{\rm c} + 2 - \sqrt{(3P_{\rm c}+2)^2 - 16P_{\rm c}}}{2} \ .
	\enq

When $P_{\rm c}$ is near 1 (corresponding to a point on the equator for the dipole case considered here) the contribution of the toroidal field to the overall structure of the poloidal field lines is small, and the above expressions serve as useful limits. The leading order terms for small $1 - P_{\rm c} \equiv \epsilon$ are found to be (cf.\ Figure \ref{fig_parameter_pc})
	\beq
	\begin{split}
	V_{\rm tor} \to & \frac{8\pi\epsilon^{3/2}}{3} + {\cal O}(\epsilon^{5/2}) \\
	H \to & \frac{32\pi s \epsilon^{5/2}}{15} + {\cal O}(\epsilon^{7/2}) \\
	\varphi_{\rm max} \to & \frac{4\sqrt{3} s \epsilon^{3/2}}{9}
	+ {\cal O}(\epsilon^{5/2})
	\end{split}
	\label{limiting_values}
	\enq

\label{lastpage}
\end{document}